\documentclass[apj,revtex4]{emulateapj}
\usepackage{rotating}
\usepackage{amsmath}
\usepackage{graphicx}
\usepackage{courier}
\begin{document}

\title{Herschel Observations and Updated Spectral Energy
Distributions of Five Sunlike Stars with Debris Disks}

\author{Sarah E. Dodson-Robinson\altaffilmark{1}, Kate Y. L.
Su\altaffilmark{2}, Geoff Bryden\altaffilmark{3}, Paul
Harvey\altaffilmark{4}, Joel D. Green\altaffilmark{4,5}}

\altaffiltext{1}{Department of Physics and Astronomy, University
of Delaware, 217 Sharp Lab, Newark, DE 19716; sdr@udel.edu}
\altaffiltext{2}{Steward Observatory, Department of Astronomy,
University of Arizona, 933 North Cherry Ave, Tucson, AZ 85721}
\altaffiltext{3}{Jet Propulsion Laboratory, California Institute
of Technology, 4800 Oak Grove Dr, Pasadena, CA 91109}
\altaffiltext{4}{Astronomy Department, University of Texas, 2515
Speedway Dr. C1400, Austin, TX 78712}
\altaffiltext{5}{Space Telescope Science Institute, 3700 San
Martin Dr, Baltimore, MD 21218}

\begin{abstract}

Observations from the {\it Herschel} Space Observatory have more than
doubled the number of wide debris disks orbiting Sunlike stars to
include over 30 systems with $R > 100$~AU. Here we present new {\it
Herschel} PACS and re-analyzed {\it Spitzer} MIPS photometry of five
Sunlike stars with wide debris disks, from Kuiper belt size to $R >
150$~AU. The disk surrounding HD~105211 is well resolved, with an
angular extent of $> 14 \arcsec$ along the major axis, and the disks of
HD~33636, HD~50554, and HD~52265 are extended beyond the PACS PSF size
(50\% of energy enclosed within radius $4 \farcs 23$). HD~105211 also
has a $24 \micron$ infrared excess that was previously overlooked
because of a poorly constrained photospheric model. Archival {\it
Spitzer} IRS observations indicate that the disks have small grains of
minimum radius $a_{\rm min} \sim 3 \micron$, though $a_{\rm min}$ is
larger than the radiation pressure blowout size in all systems. If
modeled as single-temperature blackbodies, the disk temperatures would
all be $< 60$~K. Our radiative transfer models predict actual disk radii
approximately twice the radius of a model blackbody disk. We find that
the Herschel photometry traces dust near the source population of
planetesimals. The disk luminosities are in the range $2 \times 10^{-5}
\leq L/L_{\odot} \leq 2 \times 10^{-4}$, consistent with collisions in
icy planetesimal belts stirred by Pluto-size dwarf planets.


\end{abstract}

\section{Introduction}
\label{introduction}

Observations from the {\it Herschel} Space Observatory have revealed a
population of extrasolar debris disks with blackbody temperatures $<
60$~K and peak flux densities at wavelengths near $100 \micron$.
Photodetector Array Camera and Spectrograph \citep[PACS;][]{pilbratt10}
$100 \micron$ and $160 \micron$ observations of the DUst around NEarby
Stars (DUNES) Sunlike targets reveal a debris disk incidence of 20\%, up
from 12\% at the shorter {\it Spitzer} MIPS wavelengths \citep{eiroa13}.
Far-IR observations are therefore opening up a new discovery space of
disks more than 100 times more luminous than the Edgeworth-Kuiper belt,
but too cold to detect at $70 \micron$.

We now know of over 30 disks with radii $R > 100$~AU \citep{eiroa10,
morales13, duchene14, pawellek14}. Since even wide disks require active
collisional cascades to replenish $1-10 \micron$ grains \citep{burns79},
planetesimal formation must be a robust process even at very large
distances from the star. Planet formation at 125-250~AU may even have
proceeded as far as super-Earth mass \citep{kenyon15}.  The wide dust
disks revealed by {\it Herschel} and earlier (sub)millimeter/scattered
light observations \citep[e.g.][]{greaves98, ardila04, krist05, wyatt05,
hines07, kalas07a, kalas07b, liseau08, krist10, golimowski11} may be
remnants of planetary systems that orbit Sunlike stars but are
nevertheless very different than our own: dynamical evidence indicates
that the solar nebula was likely truncated between 80 and 100~AU due to
photoevaporation or tidal interactions with nearby young stars
\citep{kretke12, anderson13}. Here we present {\it Herschel}
observations of five Sunlike stars with debris disks that transition
from Kuiper Belt size to $R > 150$~AU.  We use the Debris Disk Simulator
\citep{wolf05} to construct model SEDs and compare our disk radii, grain
sizes, and luminosities with other debris disks orbiting Sunlike stars.

We begin with a discussion of target selection, observations, and data
reduction (section \ref{sec:observations}). Next, we discuss the
extended structure observed in our images: our targets include one
well-resolved disk and three moderately extended sources (section
\ref{sec:extended}). We then present the new {\it Herschel} and {\it
Spitzer} photometry in section \ref{sec:photometry}. Radiative transfer
models of the disk spectral energy distributions (SEDs) follow in
section \ref{sec:SEDs}, with conclusions and possible future work
in section \ref{sec:conclusion}.


\section{Observations and Data Reduction}
\label{sec:observations}

We selected five Sunlike stars with debris disks discovered in {\it
Spitzer} observations for {\it Herschel} follow-up.  All targets have
infrared excess emission seen in both {\it Spitzer} IRS observations at
$32 \micron$ \citep{dodsonrobinson11, chen14} and in {\it Spitzer} MIPS
observations at $70 \micron$ \citep{beichman06, bryden09}.  (HD~105211
also has a MIPS $24 \micron$ excess, which was not realized at the time
of publication \citep{beichman06}; see section \ref{sec:HD105211} for
more detail.) The observing strategy was designed to ensure detection in
all {\it Herschel} PACS bandpasses.  Based on the two detections at
different wavelengths, \citet{dodsonrobinson11} fit single-temperature
blackbody models to the {\it Spitzer} excess emission, which we extended
to the {\it Herschel} PACS bandpasses. Exposure times were selected such
that each target would be detected at the $3 \sigma$ level, where
$\sigma$ was the quadrature sum of the instrument noise and the predicted
confusion noise reported in the Herschel Observers'
Manual\footnote{Document HERSCHEL-HSC-DOC-0876}. (In reality, some of
the sources have brighter $160 \micron$ backgrounds than predicted,
leading to detections with low significance or non-detections.)

Each observation consists of PACS medium-speed cross-scans separated in
position angle by $40^{\circ}$. Program OT1\_sdodsonr\_1 includes
simultaneous $100 \micron$/$160 \micron$ observations of all targets as
well as simultaneous $70 \micron$/$160 \micron$ observations of HD~50554
and HD~105211. Observations are scan maps with eight scan legs of length
$3 \arcmin$ and cross-scan step $4 \arcsec$. We also obtained $70
\micron$/$160 \micron$ observations of HD~50554, HD~52265, and HD~202206
from the Herschel Science Archive (program OT1\_amoromar\_1, PI A.
Moro-Mart\'{i}n). Data reduction beginning with the Level 1 data cubes was
performed using the Herschel Interactive Processing Environment version
12.1.0 \citep[HIPE;][]{ott10}. We used the HIPE implementation of the
JScanam map-making algorithm \citep{gracia15} to combine scans and
cross-scans into oversampled mosaics of $1 \arcsec$ per pixel at 70 and
$100 \micron$ and $2 \arcsec$ per pixel at $160 \micron$. The JScanam
map-maker removes the $1/f$ noise that results in signal drifts without
subtracting extended emission, which was an important consideration
because some of our sources are not pointlike (see Section
\ref{sec:extended}). However, JScanam also leaves point sources intact,
so can be used for any type of target. Table \ref{tab:targets}
(located on the last page of this document) lists the
targets in our sample and their Herschel Science Archive observation
IDs. The table also includes luminosity and age estimates from the
literature \citep{mcdonald12, chen14, bonfanti15} and luminosity
estimates from this work\footnote{We calculate luminosities based on
$V$-band magnitudes using $\frac{L_*}{L_{\odot}} =
2.512^{\left(M_{\odot} - M_* \right)}$, where the variables $M$ and $L$
represent absolute magnitude and luminosity. We find better than 10\%
agreement between our luminosity calculations and the literature values
and base all further analysis on our own luminosity calculations.}.
Figures \ref{HD33636images}, \ref{HD50554images}, \ref{HD52265images},
\ref{HD105211images}, and \ref{HD202206images} show the final mosaics in
each band for all targets.


\section{Extended Structure}
\label{sec:extended}

\begin{figure}
\centering
\includegraphics[width=0.98\textwidth]{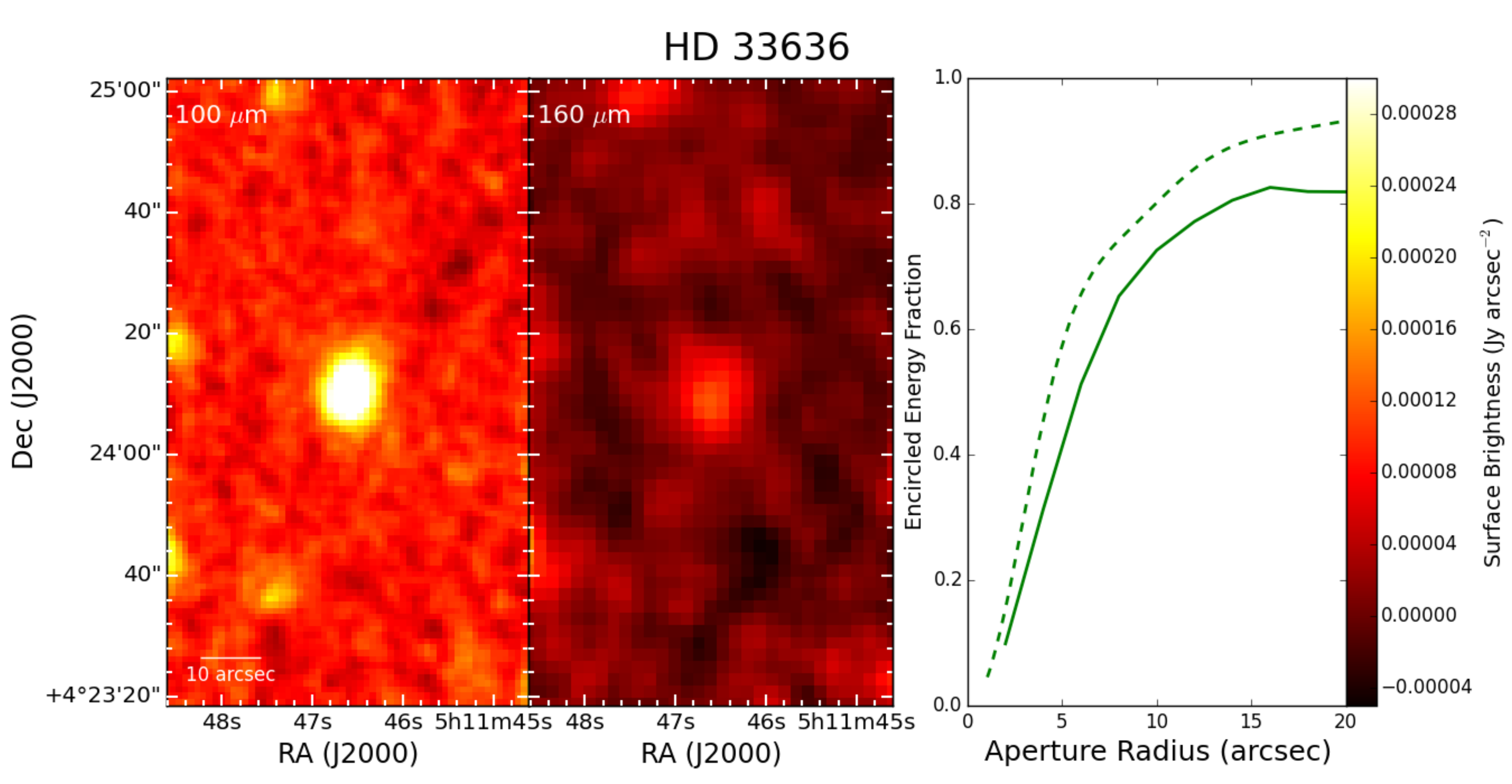}
\caption{PACS images of HD~33636. The rightmost panel shows
$100 \micron$ curves of growth for the point source Vesta (dashed
line) and HD~33636 (solid line). Encircled energy fraction rises
more slowly with aperture radius for HD~33636 than for Vesta,
indicating that HD~33636 is extended at $100 \micron$.}
\label{HD33636images}
\end{figure}

\begin{figure}
\centering
\includegraphics[width=0.98\textwidth]{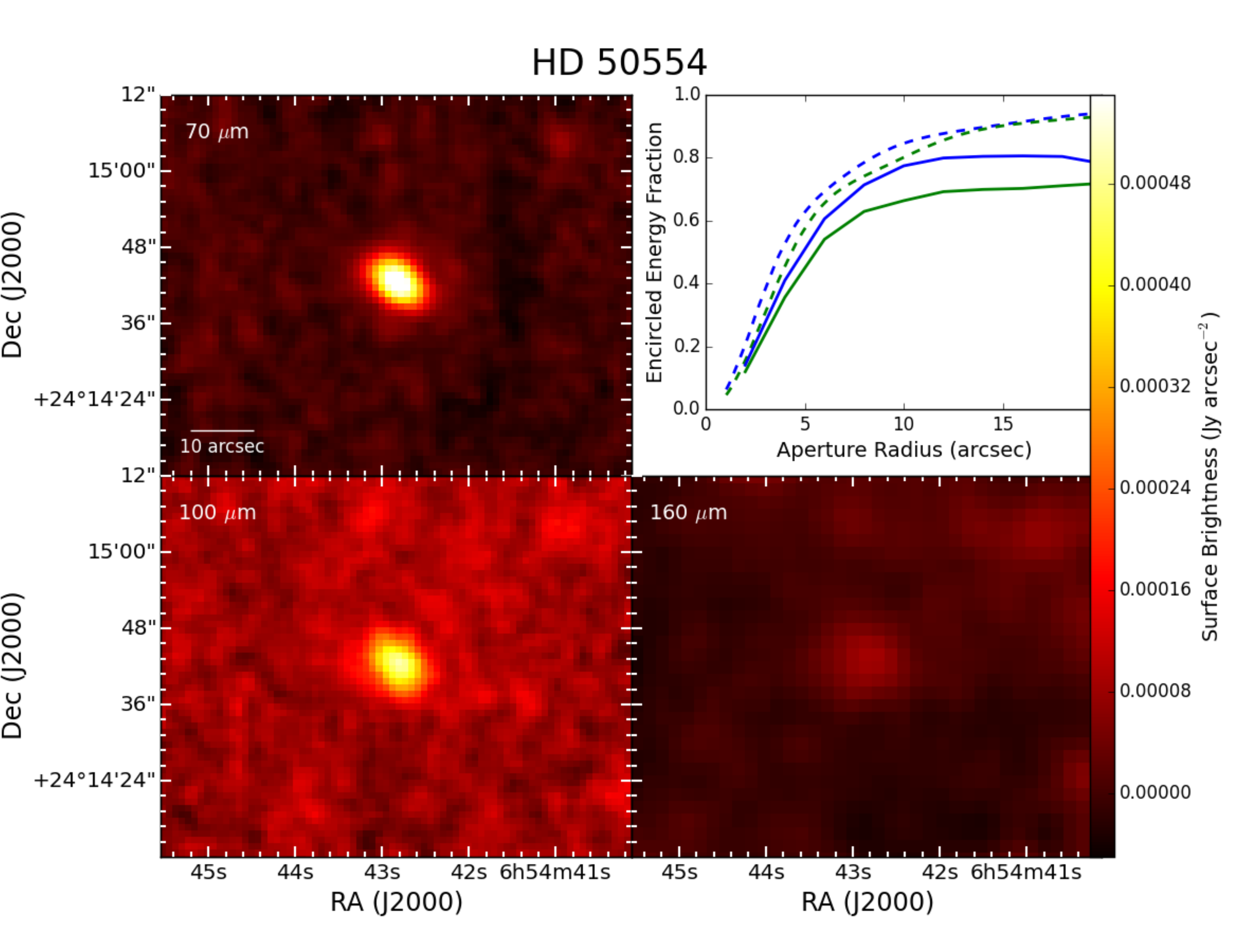}
\caption{PACS images of HD~50554. The upper-right plot shows curves of
growth for the point source Vesta (dashed lines) and HD~50554 (solid
lines), with $70 \micron$ in blue and $100 \micron$ in green. The slower
rise of HD~50554's curves of growth compared with Vesta's indicates
extended structure, which is clearly visible in the $70 \micron$ and
$100 \micron$ images.}
\label{HD50554images}
\end{figure}

\begin{figure}
\centering
\includegraphics[width=0.98\textwidth]{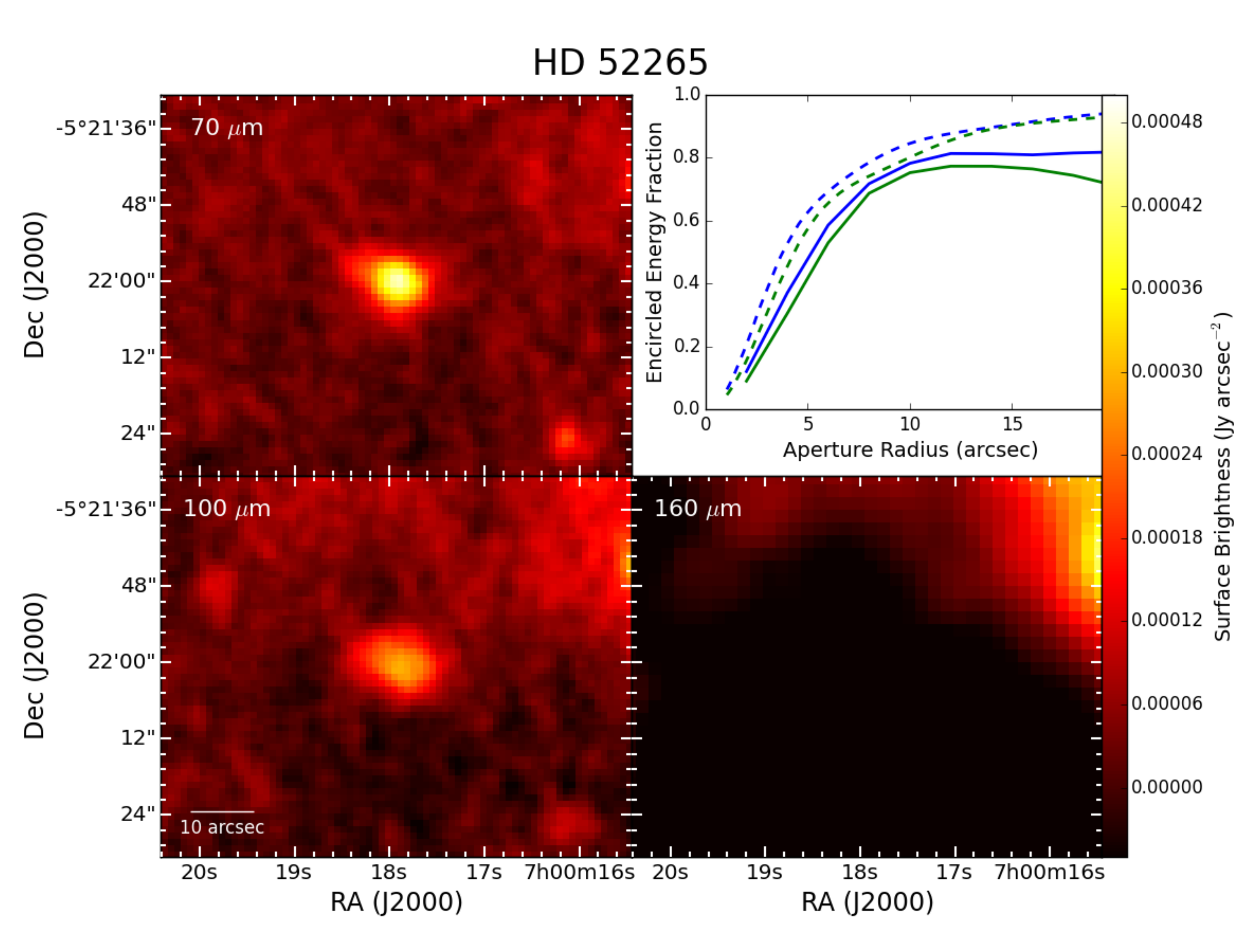}
\caption{PACS images of HD~52265. The upper-right plot shows curves of
growth for the point source Vesta (dashed lines) and HD~52265 (solid
lines), with $70 \micron$ in blue and $100 \micron$ in green. Curves of
growth suggest extended structure in the $70 \micron$ and $100 \micron$
images, though the background confusion may mimic extended structure.
Confusion noise prevents detection of HD~52265 at $160 \micron$.}
\label{HD52265images}
\end{figure}

\begin{figure}
\centering
\includegraphics[width=0.98\textwidth]{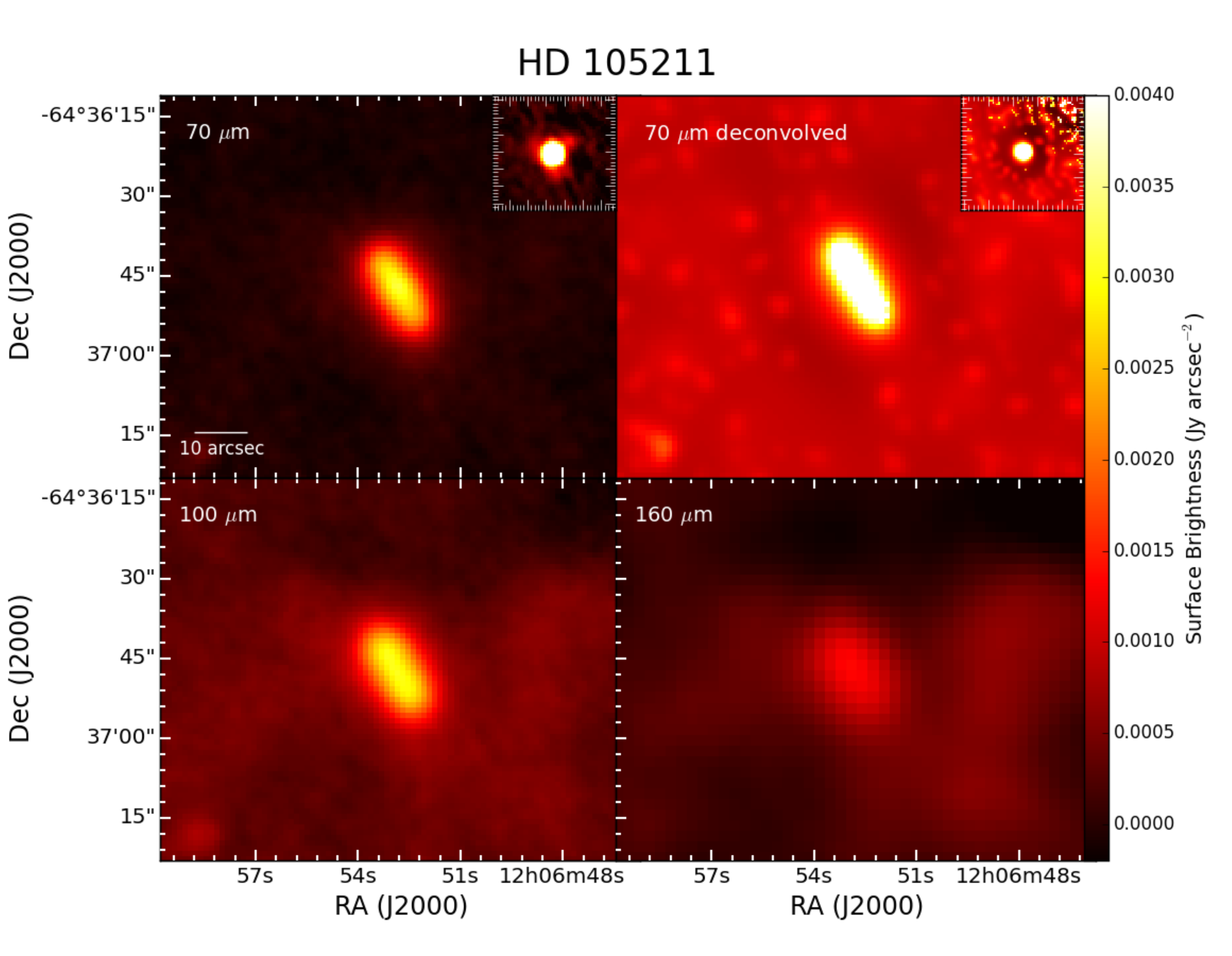}
\caption{PACS images of HD~105211. The disk is well resolved at all PACS
wavelengths. The deconvolved $70 \micron$ image in the upper-right panel
was created using the PSF measured from a point source in the
$70^{\circ}$ cross-scan. The insets in the $70 \micron$ plot panels show
the point source before deconvolution (left), with the tri-lobe
structure visible, and after deconvolution (right).}

\label{HD105211images}
\end{figure}

\begin{figure}
\centering
\includegraphics[width=0.98\textwidth]{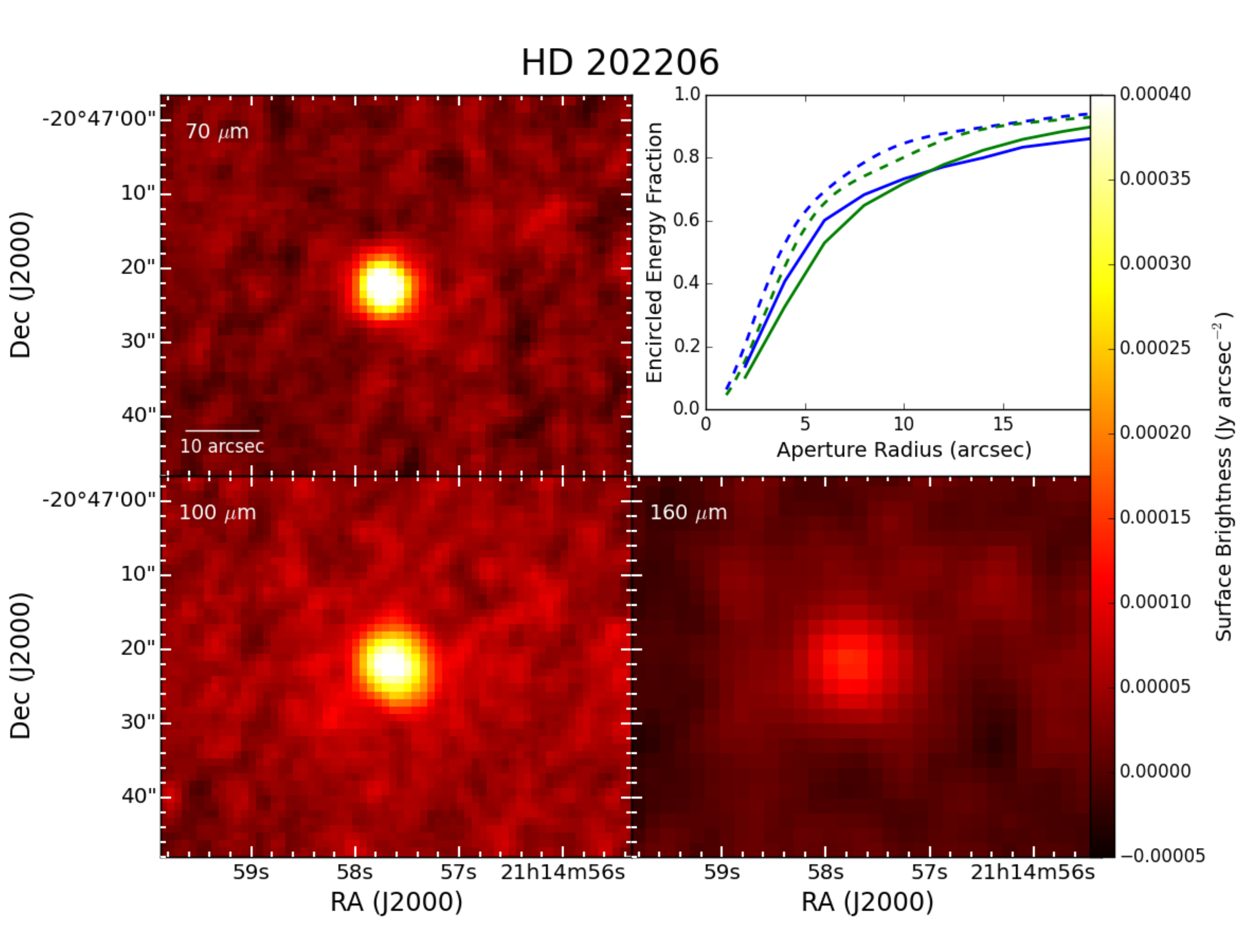}
\caption{PACS images of HD~202206. $70 \micron$ and $100 \micron$
curves of growth do not conclusively show extended structure.}
\label{HD202206images}
\end{figure}

Determining the spatial extent of the disks is important for selecting
the correct aperture sizes for photometry. Examining the $70 \micron$
and $100 \micron$ mosaics in Figures \ref{HD33636images},
\ref{HD50554images}, \ref{HD52265images}, \ref{HD105211images} and
\ref{HD202206images} suggests that the disks are not point sources, but
instead have some extended structure. Figures \ref{HD33636images},
\ref{HD50554images}, \ref{HD52265images}, and \ref{HD202206images}
contain curves of growth for HD~33636, HD~50554, HD~52265, and HD~202206
showing encircled energy fraction as a function of circular aperture
radius for the $100 \micron$ (green) and, where available, $70 \micron$
(blue) mosaics. For comparison, the dashed lines show curves of growth
from the point source Vesta \citep{lutz10}. For HD~33636, HD~50554, and
HD~52265, the curves of growth are well below those of the Vesta PSF,
indicating that the sources have extended structure. HD~50554 is
elongated at both $70 \micron$ and $100 \micron$, but has a small
apparent position angle shift between the two bands. Fitting an
elliptical Gaussian function to the source, we find a position angle
(East of North) of $54^{\circ}$ at $70 \micron$ and $51^{\circ}$ at $100
\micron$. The $3^{\circ}$ difference is too small to conclusively
indicate morphology changes between dust emitting at different
wavelengths, but should be kept in mind as a possible source of
uncertainty in the photometry and SED fitting. HD~202206 is less well
resolved than the other sources, but is still more extended than a point
source.  For some targets some cases the curves of growth turn downward
instead of monotonically increasing with aperture size, indicating that
the aperture has expanded to include pixels with negative flux
densities. The photometric aperture must be kept smaller than the
curve-of-growth turnover radius. HD~105211 (Figure \ref{HD105211images})
is clearly resolved in all PACS bands. The extended nature of our
targets leads us use larger photometric apertures than the faint-source
sizes used by the DUNES team \citep{eiroa13}. We use $12 \arcsec$ as our
nominal aperture size at 70 and $100 \micron$ (see Section
\ref{sec:photometry}), which captures the extended emission yet avoids
negative-valued pixels.

The upper-right panel of Figure \ref{HD105211images} shows the
deconvolved $70 \micron$ image of HD~105211. Examination of a point
source located in the 70$^{\circ}$ cross-scan, shown in the inset plots
of the original image (upper left) and the deconvolved image, reveals
the extent to which deconvolution has suppressed the trefoil PSF
structure.  Although our deconvolution procedure does not perfectly
conserve flux, one can still see that most of the emission in the
deconvolved image is concentrated in a narrow band indicating a nearly
edge-on disk. We performed an elliptical Gaussian source fit on the
original (not deconvolved) $70 \micron$ image and found $\sigma_x = 6.8
\arcsec$ and $\sigma_y = 3.9 \arcsec$, where the x-direction is
East-West and the y-direction is North-South. The position angle is
$29.7^{\circ}$ East of North. Assuming a circular debris ring, the
radius of the $70 \micron$-emitting ring is approximately
$\sqrt{\sigma_x^2 + \sigma_y^2} = 7.9 \arcsec$, or 154~AU, in rough
agreement with the SED fitting results (see section \ref{sec:SEDs}). The
inner edge of the ring is not resolved, possibly due to the high
inclination of the disk.

\section{Photometry}
\label{sec:photometry}


For all targets except for HD~105211, our aperture photometry is based
on the PACS \texttt{ipipe} script
\texttt{L3\_pointSourceAperturePhotometry.py}\footnote{Author: Markus
Nielbock; available in HIPE through the `Scripts $\rightarrow$ PACS Useful
Scripts' menu},\\ which uses the HIPE tasks
\texttt{annularSkyAperturePhotometry} and \\
\texttt{photApertureCorrectionPointSource}.  For HD~33636, HD~50554, and
HD~202206, aperture sizes are $12 \arcsec$ at $70 \micron$ and $100
\micron$, with background subtraction based on sky annuli of $25-35
\arcsec$. For HD~52265 we use an $8 \arcsec$ aperture and an $18-22
\arcsec$ sky annulus at both $70 \micron$ and $100 \micron$ to exclude a
bright source west of the target. At $160 \micron$, where confusion
noise is most severe, we use a $14 \arcsec$ aperture and decrease the
sky annulus to $24-28 \arcsec$ for HD~33636 and HD~202206.  For HD~50554
at $160 \micron$ we use an aperture of $10 \arcsec$ and a sky annulus of
$20-24 \arcsec$ in order to avoid a nearby source. HD~52265 is
undetected at $160 \micron$ due to confusion.  For the extended source
HD~105211, we use the Aperture Photometry Tool \citep[APT;][]{laher12}.
We choose an elliptical aperture of semimajor axis $18 \arcsec$ and
semiminor axis $12 \arcsec$ at an angle of $30^{\circ}$ East of North.
The sky annulus has major axis $30-45 \arcsec$ and minor axis $20-30
\arcsec$ at 70 and $100 \micron$, and major axis $30-44 \arcsec$ and
minor axis $20-29.3 \arcsec$ at $160 \micron$. The elliptical aperture
corrections are based on Table 2 of \citet{balog14}, which reports
encircled energy fraction (EEF) as a function of circular aperture
radius for each PACS filter. For an aperture with semimajor axis $a$ and
semiminor axis $b$, the multiplicative aperture correction $A$ is
\begin{equation} A = \frac{1}{\sqrt{{\rm EEF} \left( a \right) \times
{{\rm EEF} \left( b \right)}}}.  \label{eq:apcorrect} \end{equation} For
$a = 18 \arcsec$ and $b = 12 \arcsec$, we find aperture corrections of
1.201 at $70 \micron$, 1.241 at $100 \micron$, and 1.384 at $160
\micron$.

Uncertainties have several components: Poisson noise from the source,
sky background, PACS absolute flux calibration, and residual $1/f$ noise
that may elude the map-making algorithm. Since the pixels in each final
mosaic are not native, but instead are reconstructed from multiple scans
of the source, they are not independent and the method of estimating
random noise as $\sigma_{\rm sky} \sqrt{N_{\rm pix}}$ (where $\sigma$ is
the standard deviation of the sky flux and $N_{\rm pix}$ is the number
of pixels in the aperture) does not work. Instead, the usual procedure
for calculating the random component of the PACS-image error budget is
to place apertures of the same size and shape used for the science
target on the sky background and calculate the standard deviation of the
aperture-corrected flux from each. Following \citet{balog14}, we use six
apertures placed evenly around the outer edge of the background sky
annulus for all targets except HD~105211. The HD~105211 sky background
is structured, with an extended bright source directly to the west of
the disk (Figure \ref{HD105211images}; see especially the $160 \micron$
image at the lower right). To estimate photometric error for HD~105211,
we use 12 elliptical apertures placed randomly on the image, excluding
the bright source. Finally, we include conservative estimates of the
systematic errors---both from absolute flux calibration and
repeatability---of 7\% of the source flux \citep{balog14}. Quoted
uncertainties are the quadrature sum of the random and systematic
errors. Table \ref{tab:photometry} lists the measured PACS flux
densities.

As a consistency check, we compare our PACS $70 \micron$ flux densities
with {\it Spitzer} MIPS $70 \micron$ observations. Published values come
from the catalog of \citet{chen14}, which includes photometry from a
variety of literature sources. We also present new photometry of both
the $24 \micron$ and $70 \micron$ images based on the MIPS team in-house
pipeline \citep{gordon05, su10}\footnote{New photometry presented in
Table \ref{tab:photometry} is labeled ``\citet{su10}'' under Analysis
Method.}. MIPS did not observe HD~202206 at $24 \micron$. We use PSF
extraction to find $24 \micron$ flux densities for the remaining
targets. All targets have MIPS $70 \micron$ data. We use PSF extraction
to measure $70 \micron$ flux densities for all targets except HD~105211,
for which we use an aperture with radius $35 \arcsec$ and a sky annulus
of $39-65 \arcsec$.

Table \ref{tab:photometry} lists the MIPS flux densities for each
target. The two MIPS $24 \micron$ analysis pipelines yield flux
densities that are consistent at the 1--2$\sigma$ level, though there is
a clear discrepancy in how the error bars are computed for HD~105211.
The MIPS $70 \micron$ flux densities from the two pipelines are fully
consistent for each source except HD~105211. The low HD~105211 $70
\micron$ flux density of 474~mJy reported by \citet{beichman06} is
probably due to their choice to use a uniform aperture radius of 1.5
camera pixels ($7 \farcs 8$) for the entire survey. Since the spatial
extent of the HD~105211 disk in our PACS $70 \micron$ images is $> 14
\arcsec$ on the major axis, \citet{beichman06} may have underestimated
the flux density. The remaining PACS and MIPS $70 \micron$ fluxes are
consistent at the $1\sigma$ level, indicating good agreement between
the {\it Spitzer} and {\it Herschel} fluxes.

As a final test of our photometry pipeline, we measure fluxes from the
PACS calibrator star $\alpha$~Ceti (HD 18884). We use scan/cross-scan
AOR pairs 1342212853/1342212854 ($100/160 \micron$; observing day 614)
and 1342203030/1342203031 ($70/160 \micron$; observing day 457). The
scan/cross-scan pairs were processed with the same JScanam map-making
script as the science observations. We then performed photometry with
the nominal aperture and sky annuli used for HD~33636, HD~52265, and
HD~202206 (our compact, uncrowded targets). Our results match those of
the instrument team \citep[Table 12][]{balog14} to within 1\% at all
wavelengths, demonstrating the high accuracy of our data pipeline for
pointlike sources with clean backgrounds (see Section \ref{sec:SEDs} for
caveats about the HD~105211 observations).

In the next section we combine {\it Spitzer} and {\it Herschel}
observations to extend SED coverage to the peak wavelengths of dust
emission, using a radiative transfer model to derive dust properties.

\section{Spectral Energy Distributions}
\label{sec:SEDs}

The disks in this study are cold enough that observations at $70
\micron$ and shorter wavelengths probe only the increasing side of the
$F_{\nu}(\lambda)$ curve. The {\it Herschel} photometry extends coverage
to the peak and the decreasing side of $F_{\nu}(\lambda)$. We use the
Debris Disk Simulator (DDS)\footnote{www1.astrophysik.uni-kiel.de/dds/}
\citep{wolf03, wolf05} to fit SED models to the {\it Spitzer} IRS/MIPS
and {\it Herschel} PACS photometry. The DDS takes as input a stellar
spectrum, an inner and outer radius for the dust ring, a dust mass, an
analytical or numerical description of the dust number density $n(r)$
(where $r$ is the distance from the star), and a grain size distribution
of the form $n(a) \propto a^{-q} ; a_{\rm min} \leq a \leq a_{\rm max}$
($a$ is the grain radius and we set $q = 3.5$). The assumption of a
power-law grain size distribution is reasonable for the larger grains,
but dynamical models that include loss processes such as
Poynting-Robertson drag predict that for grains with $a \la 10 \micron$,
the size distribution turns over so that the grain abundance {\it
increases} with radius \citep{wyatt11, kenyon16}. Since the far
infrared-emitting dust grains are most likely part of a collisional
cascade that includes large grains, we set $a_{\rm max} = 1$~mm for all
models. Larger particles would be present, but the grain size
distribution is so steep that the larger grains would contribute only
$\left( a_{\rm max} / 1 \: \micron \right)^{-1/2} = 3\%$ of the total
thermal emission. The DDS also allows the user to set the chemical
composition of the grains; here we assume 100\% astronomical silicate
\citep{laor93}. Since we do not explore the composition options, our
model SEDs are plausible but not unique. In particular, the grains could
have organic components or ice mantles, though ice would eventually
succumb to photodesorption. However, the small number of spectral data
points that cover the bulk of the disks' emission do not warrant
extremely flexible fits.

\begin{figure}
\centering
\includegraphics[width=0.98\textwidth]{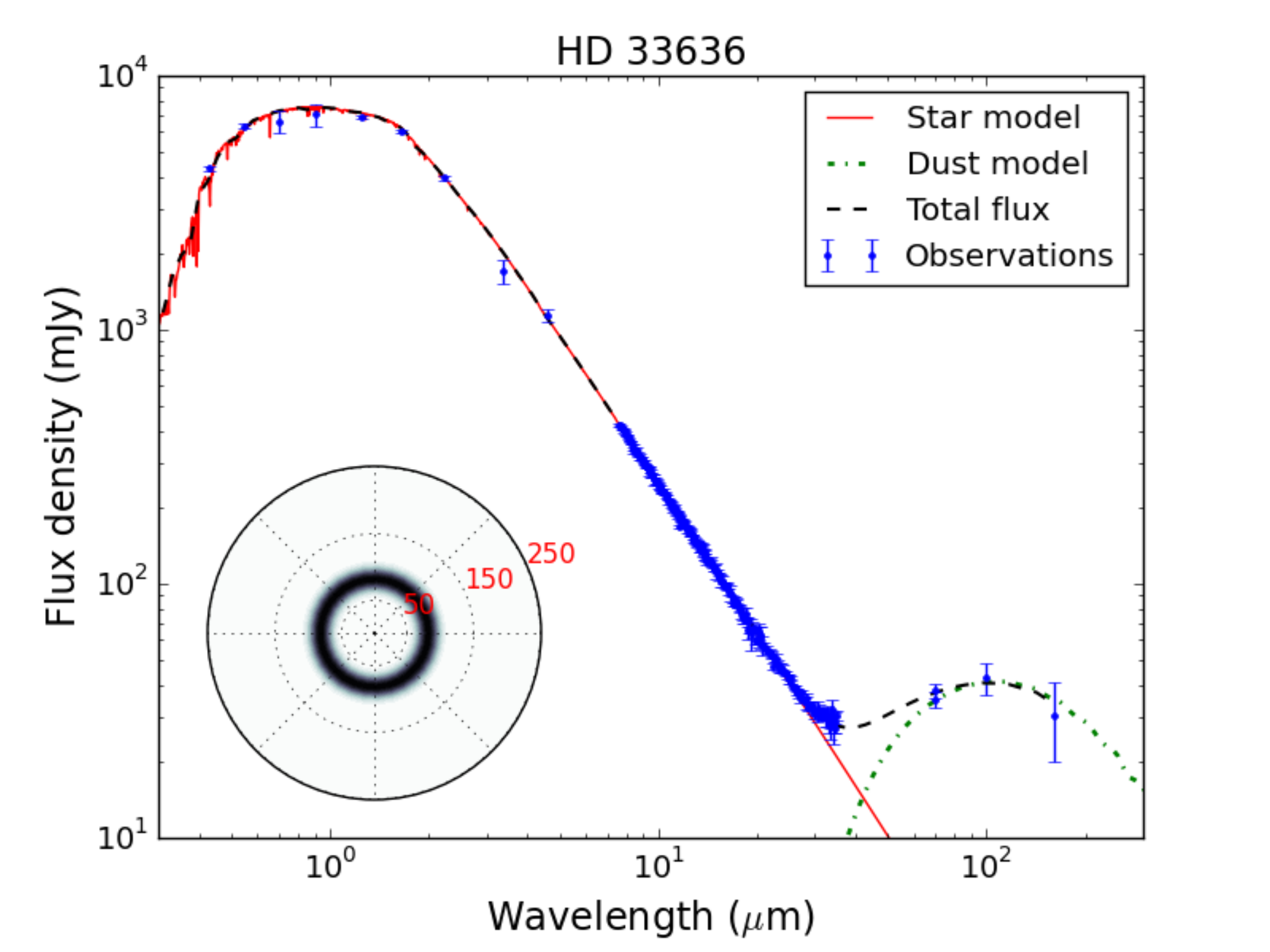} \caption{Spectral
energy distribution of the HD~33636 star-disk system.  The inset plot at
the lower left shows the best-fit dust ring model, which is centered at
85~AU and has width 12~AU. Observational data include Hipparcos $BV$,
$RI$ from \citet{bessell90}, 2MASS $JHK$, allWISE, {\it Spitzer} IRS and
MIPS, and PACS 100/$160 \micron$.}
\label{HD33636sed}
\end{figure}
\begin{figure}
\centering
\includegraphics[width=0.98\textwidth]{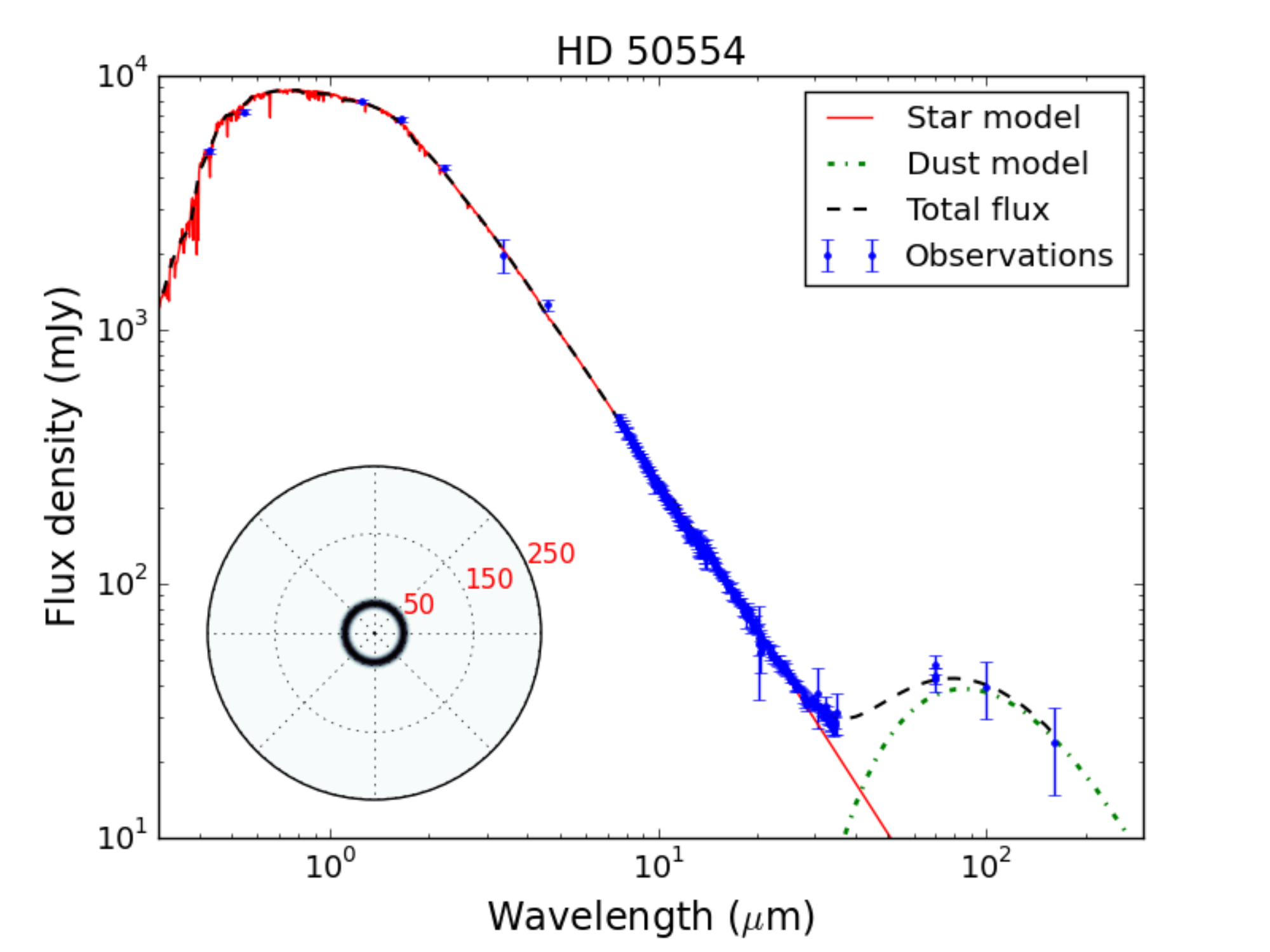}
\caption{Spectral energy distribution of the HD~50554 star-disk system.
The lower-left inset shows the best-fit Gaussian ring model with center
45~AU and width 4~AU. Observational data include Hipparcos $BV$,
2MASS $JHK$, allWISE, {\it Spitzer} IRS and MIPS, and {\it Herschel}
PACS.}
\label{HD50554sed}
\end{figure}
\begin{figure}
\centering
\includegraphics[width=0.98\textwidth]{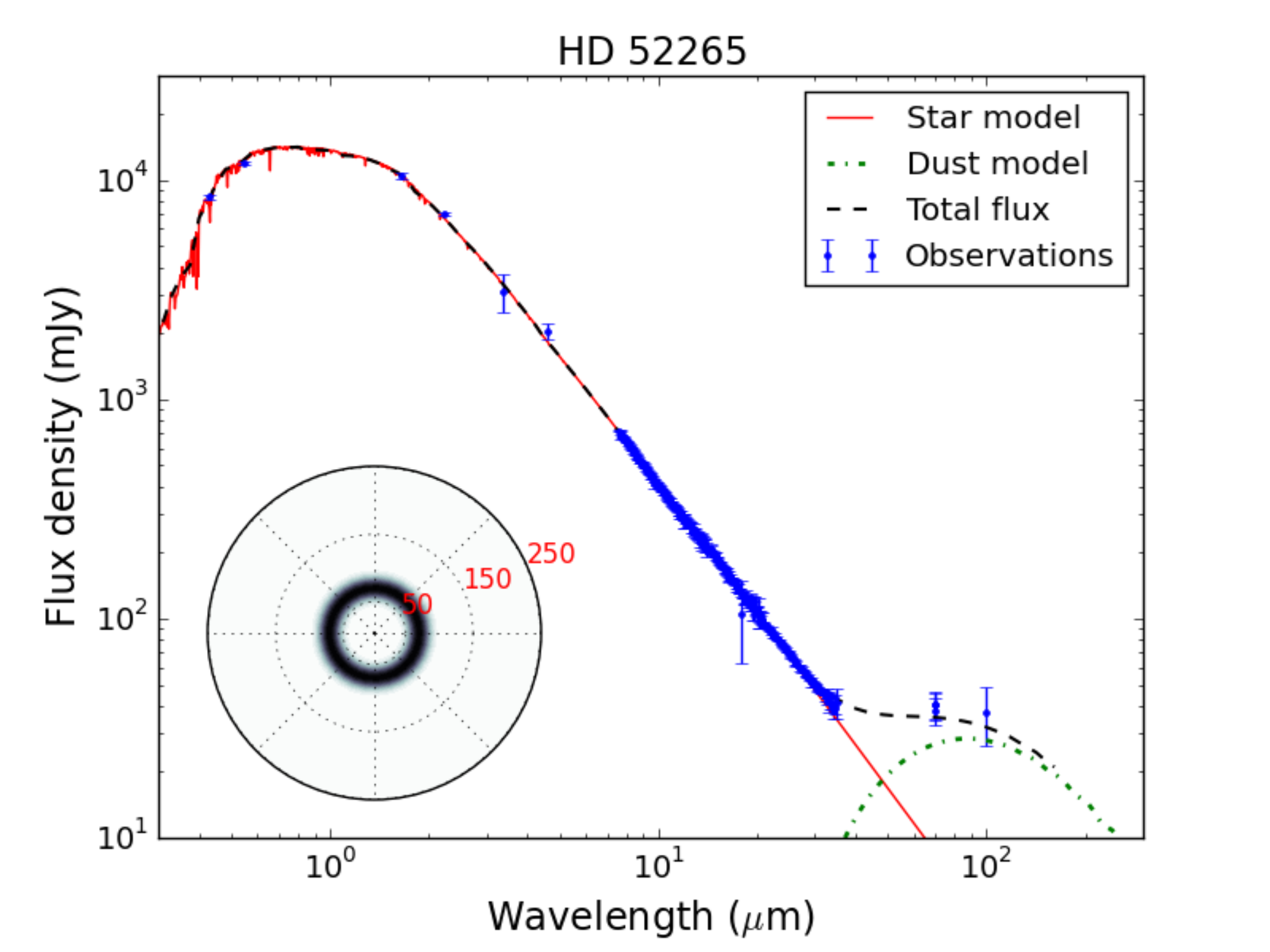}
\caption{Spectral energy distribution of the HD~52265 star-disk
system. The lower-left inset shows the best-fit Gaussian ring
model with center 70~AU and width 10~AU. Observational data
include Hipparcos $BV$, 2MASS $HK$, allWISE, {\it Spitzer} IRS
and MIPS, and {\it Herschel} PACS. HD~52265 was not detected at
$160 \micron$. 2MASS $J$ photometry was flagged as low quality
and was not included in the analysis.}
\label{HD52265sed}
\end{figure}

The parameters we explore are minimum grain size $a_{\rm min}$, the
debris ring's distance from the star $R$, the width of the debris ring
$\sigma$ (assuming a Gaussian number-density profile centered at $R$),
and the mass of grains smaller than $a_{\rm max} = 1$~mm.  Input star
spectra are Atlas9 model photospheres \citep{kurucz92} originally fitted
by \citet{dodsonrobinson11}, based on {\it Hipparcos/Tycho} $BT$ and
$VT$ magnitudes \citep{perryman97, hog00} and 2MASS $JHK$ magnitudes
\citep{cutri03}. Where available, we added $RI$ photometry from
\citet{bessell90}. We also checked for consistency between the
\citet{dodsonrobinson11} model photospheres and the allWISE photometry
\citep{wright10, cutri13} and found good agreement. The allWISE
photometry was particularly important for HD~105211, which has
conflicting photosphere models in the literature due to poor-quality
2MASS photometry; and for HD~52265, which has a poor-quality 2MASS $J$
magnitude. Table \ref{tab:seds} lists the fitted parameters for each
SED, and Figures \ref{HD33636sed}-\ref{HD202206sed} show the model
SEDs. We begin our discussion of SEDs with a detailed look at the
photosphere model and mid-IR photometry of HD~105211, which requires
extra care because of the imprecise 2MASS data, then move on to the
ensemble properties of our disks.

\subsection{HD 105211 Mid-IR Photometry and Photosphere}
\label{sec:HD105211}

\begin{figure}
\centering
\includegraphics[width=0.98\textwidth]{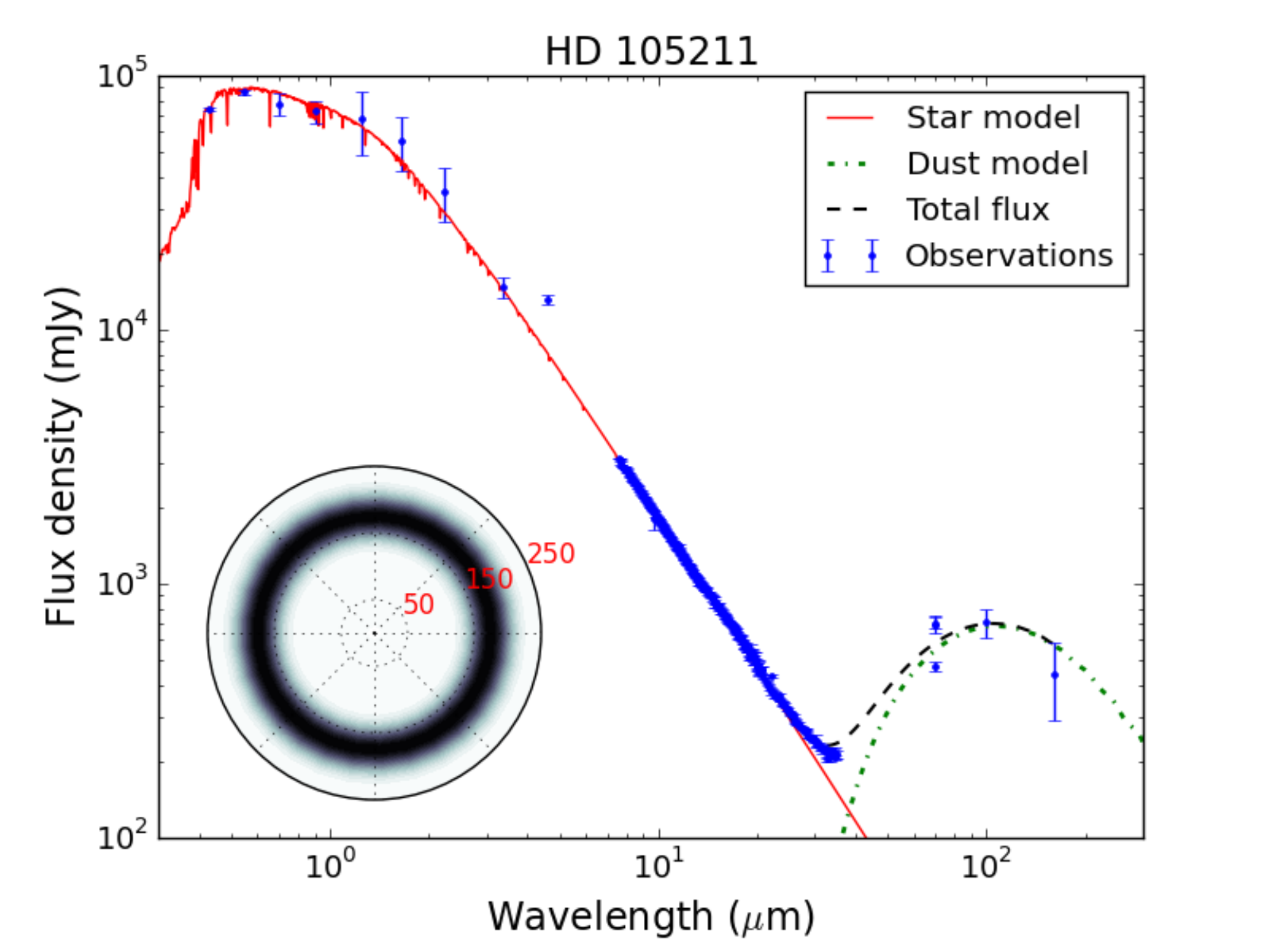}
\caption{Spectral energy distribution of the HD~105211 star-disk
system. The lower-left inset shows the best-fit Gaussian ring
model with center 175~AU and width 20~AU. Observational data
include Hipparcos $BV$, \citet{bessell90} $RI$, 2MASS $JHK$,
allWISE, {\it Spitzer} IRS and MIPS, and {\it Herschel} PACS.
HD~105211 is saturated in the allWISE W2 filter.}
\label{HD105211sed}
\end{figure}

\begin{figure}
\centering
\includegraphics[width=0.98\textwidth]{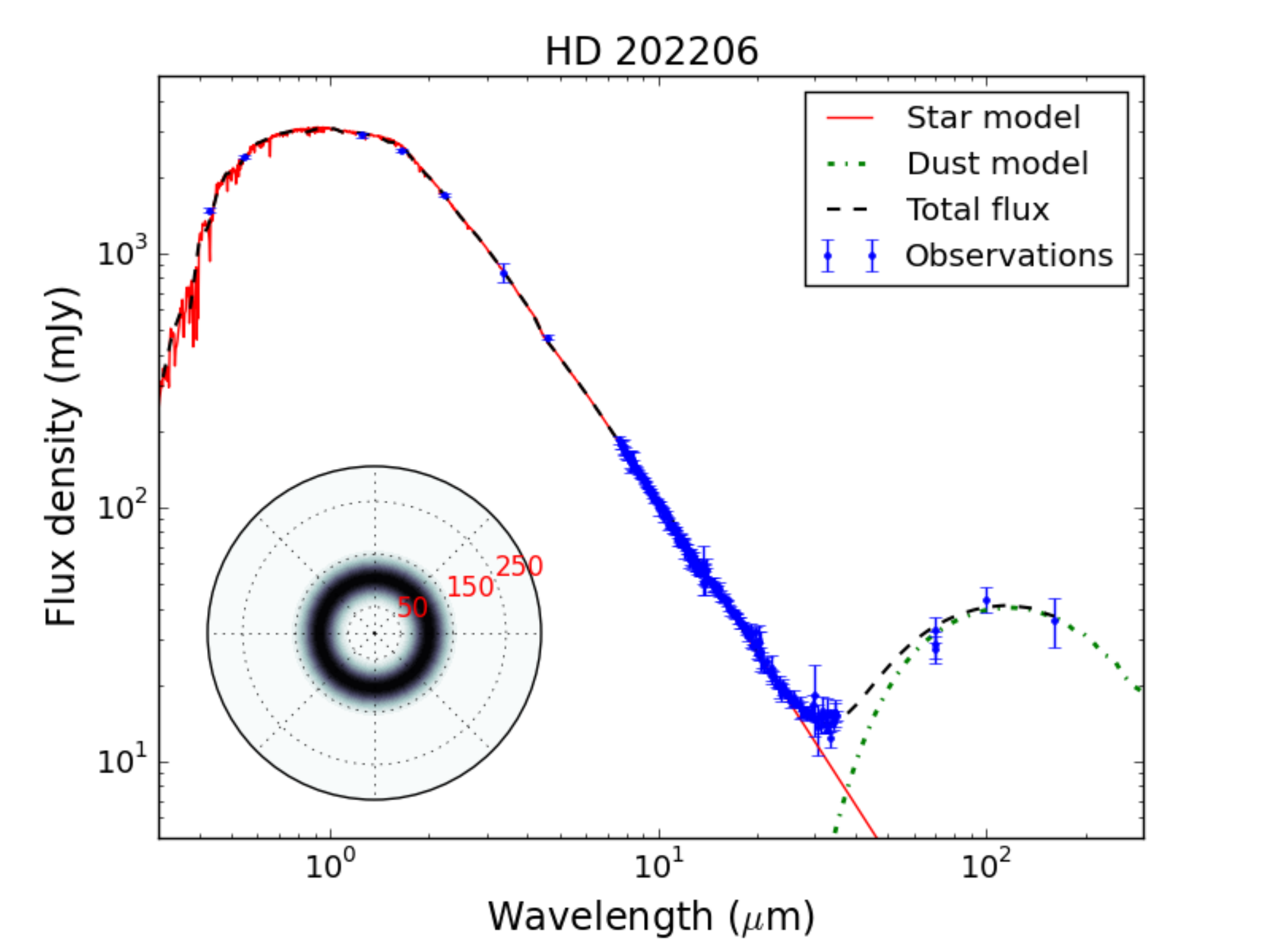}
\caption{Spectral energy distribution of the HD~202205 star-disk
system. The lower-left inset shows the best-fit Gaussian ring
model with center 105~AU and width 20~AU. Observational data
include Hipparcos $BV$, 2MASS $JHK$, allWISE, {\it Spitzer} IRS
and MIPS, and {\it Herschel PACS}.}
\label{HD202206sed}
\end{figure}

HD~105211 is bright enough to have saturated the 2MASS detectors, so the
$JHK$ flux densities have $> 20$\% errors. The near-infrared data is
important for constraining the Rayleigh-Jeans side of the spectrum, as
the turnover to an $F_{\nu} \propto \lambda^{-2}$ power law happens in
the $H$ band for HD~105211. A further issue is that the SIMBAD database
identifies HD~105211 as a spectroscopic binary, though
\citet{eggleton08} find that a single star is the most probable
configuration. If HD~105211 is a binary, a single photosphere model
might not accurately reproduce the emission from both components.
\citet{beichman06} presented the first {\it Spitzer} observations of
HD~105211, reporting an infrared excess at $70 \micron$ but not at $24
\micron$. Their measured excess was based on a model photosphere with
$T_{\rm eff} = 6600$~K, which was fit to $BVJHK$ photometry and
predicted a photospheric flux density of 363.4~mJy at $24 \micron$.
\citet{saffe08} then measured $T_{\rm eff} = 6900$~K from spectroscopy,
which agrees with the optical spectral type F2~V of
\citet{gray06}. \citet{dodsonrobinson11} added $RI$ photometry from
\citet{bessell90} to the photospheric fit constraints, which yielded a
best-fit photosphere model with $T_{\rm eff} = 7250$~K.  Moving to the
new, hotter photosphere model yields a higher near-IR/mid-IR flux ratio
and a lower predicted photospheric $24 \micron$ flux density of
321.9~mJy.  Based on the \citet{dodsonrobinson11} model photosphere and
our re-reduction of the MIPS data, HD~105211 has a $13 \sigma$ excess at
$24 \micron$ (see Table \ref{tab:photometry}). \citet{chen14} computed a
third model photosphere with a predicted $24 \micron$ flux of 347.1~mJy,
which combined with the new MIPS re-reduction yields a $7\sigma$
excess at $24 \micron$.
The allWISE W1 photometry best corresponds to the
\citet{dodsonrobinson11} photosphere model, with a monochromatic $3.35
\micron$ flux density only $0.11 \sigma$ higher than the model. As the
star is saturated in W2 (see Figure \ref{HD105211sed}), we do not re-fit
the model photosphere with the allWISE data added. We adopt the
\citet{dodsonrobinson11} photosphere model for this work, but note
that the accuracy of the spectral type and effective temperature would
benefit from additional data. \citet{james13} find an intrinsic scatter
of about two spectral subtypes in a cross-correlation analysis of their
spectral library---for example, the spectrum of a star typed F8~V might
best match the F6~V template spectrum---a level of uncertainty
consistent with the differences between the analyses of
\citet{dodsonrobinson11}, \citet{chen14}, \citet{saffe08}, and
\citet{gray06}.

There is still concern about the IRS data, however. Since slit loss led
to 5-10\% flux density underestimates in IRS, \citet{dodsonrobinson11}
calculated absolute flux densities by assuming that the
shortest-wavelength data points of the SL1 module (beginning at $7.58
\micron$) traced the photosphere. The LL2 and LL1 data were then spliced
to the SL1 data so as to make a continuous function. However, the data
at the long-wavelength end of LL2 ($19-20 \micron$) have large error
bars, and there appears to be a drop in fractional excess flux at the
short-wavelength end of LL1 ($> 20 \micron$; see Figure 2 of
\citet{dodsonrobinson11}). It is possible that HD~105211 has multiple
debris belts, which would account for the rise and fall of fractional
excess at $19-20 \micron$, but the IRS data are not conclusive. What is
concerning is that the IRS data differ from the allWISE W4 data (after
conversion to monochromatic flux at $22.1 \micron$) by $\sim 5 \sigma$.
(\citet{beichman06} and our MIPS re-reduction disagree on the errors in
the MIPS $24 \micron$ flux, but adopting $\sigma = 22.5$~mJy from
\citet{beichman06}, our IRS $24 \micron$ flux density is consistent with
the MIPS data.) The disagreement between W4 and IRS may be due to the W4
color correction; the bandpass is broad and the SED is no longer
photospheric. The IRS, allWISE, and MIPS data all indicate an infrared
excess at $22-24 \micron$, but the level of excess is still open to
debate.

We also have some concerns about the photometry at $160 \micron$ due to
the bright source east of the target, which may have led to
over-subtraction of the sky background.  We find it difficult to
reproduce the ``peakiness'' of the SED at $100 \micron$; dust models
that fit the $32/70 \micron$ flux ratio tend to over-predict the $160
\micron$ flux density and models that reproduce the $70/100 \micron$
flux ratio tend to under-predict the $32 \micron$ flux density.  A
low abundance of grains near the minimum size, as suggested by
collisional cascade models that predict non power-law size distributions
\citep{wyatt11, kenyon16}, might improve the SED fit by decreasing the
$32/70 \micron$ flux ratio. The SED model shown in Figure
\ref{HD105211sed} is a compromise between the opposing constraints at 32
and $70 \micron$. Reducing the maximum grain size that contributes to
the emission from the 1~mm assumed for all disks could reduce the
predicted $160 \micron$ flux density, but we doubt the physical realism
of any model that includes small grains without a continuous size
distribution extending to macroscopic grains.  The uncertainties in the
PACS and MIPS photometry (section \ref{sec:photometry}) and the
underlying photosphere model must be kept in mind when discussing the
dust properties. Based on the infrared excess at both $24 \micron$ and
$70-160 \micron$, we strongly suggest follow-up observations of
HD~105211 with both JWST and ALMA.


\subsection{Dust Properties}
\label{sec:dust}

We now turn our attention to the properties of the dust in our systems,
including the disk radius, minimum grain size, location of dust relative
to the parent planetesimal population, and possibility of planet
sculpting. First, we find that all of the dust SEDs peak between $80
\micron$ and $120 \micron$, which is typical of debris disks from the
DUNES program \citep{krivov13}. We do not find any sources that are
brighter at $160 \micron$ than $100 \micron$, a feature which could
indicate either a background galaxy in the beam \citep{gaspar14} or a
disk of unstirred, primordial grains \citep{krivov13}. We diagnose the
presence of superheated small grains by computing the parameter $\Gamma
= R / R_{\rm BB}$, the ratio of the disk radius from our SED models (in
which dust temperature depends on grain size) to the disk radius if only
blackbody grains were present:
\begin{align}
T_{\rm BB} = 5100 \: {\rm K} (1 \; \micron/\lambda_{\rm max}) \\
R_{\rm BB} = \left( \frac{278 \; {\rm K}}{T_{\rm BB}} \right)^2
\left( \frac{L_*}{L_{\odot}} \right)^{1/2} \; {\rm AU}.
\label{eq:blackbody}
\end{align}
In equation \ref{eq:blackbody}, $T_{\rm BB}$ is the temperature of
blackbody grains, $\lambda_{\rm max}$ is the wavelength of maximum
emission, and $R_{\rm BB}$ is the radius of the ring of blackbody dust.
The disks in our sample have $1.7 < \Gamma < 2.7$ (Table
\ref{tab:seds}).  Although four out of five of our disks are not well
enough resolved to measure disk sizes directly from the images, our
SED-based values of $\Gamma$ agree well with studies of resolved debris
disks \citep[e.g.][]{rodriguez12, booth13}. Each disk must contain a
population of small grains that do not emit efficiently at $\sim 100
\micron$ wavelengths and so heat above the local blackbody temperature.
Given the presence of small grains, our sources are most likely
self-stirred or planet-stirred planetesimal belts with grain-producing
collisional cascades, which is consistent with our assumption of $q =
3.5$ \citep{pan12, matthews14}.

\citet{dodsonrobinson11} estimated blackbody temperatures $T_{\rm BB}$
for our target stars based on {\it Spitzer} IRS and MIPS
(spectro)photometry at $32 \micron$ and $70 \micron$. For the warmer
debris systems HD~50554, HD~52265, and 105211, their estimates agree
well with ours. For HD~33636 and HD~202206, the peak flux density is
much redder than $70 \micron$ and the {\it Herschel} observations
indicate colder blackbody temperatures than the {\it Spitzer}-based
photometry. We emphasize that the SEDs modeled here are not based on
blackbody grains---our comparison between single-temperature blackbody
models simply highlights the importance of far-IR photometry in
characterizing the debris rings.

The {\it Spitzer} IRS spectra provide the best constraint on minimum
grain size. In the Rayleigh regime, where $2 \pi a \ll \lambda$,
the grain opacity $\kappa$ declines with wavelength as $\kappa \propto
\lambda^{-\beta}$, where $\beta = 2$ for simple conductors and
insulators \citep{draine06}. At $32 \micron$, the center of the spectral
window used by \citet{dodsonrobinson11} to detect infrared excesses, $2
\pi a = \lambda$ for $\lambda = 5.1 \micron$. We find that the lack of
{\it Spitzer} IRS fractional excesses at $20 \micron$ (with the
exception of HD~105211) and the small, though detectable, excesses at
$32 \micron$ exclude grains smaller than $\sim 3-4 \micron$, as such
grains would emit efficiently from $20-30 \micron$. The minimum grain
sizes predicted by our best-fit model SEDs are in good agreement with
the \citet{pawellek15} SED-based measurements from 32 objects (see their
Figure 4): all of their target stars with $L_{\odot} < L_* \leq 10
L_{\odot}$ have $\sim 2 \micron < a_{\rm min} < \sim 10 \micron$.

There is, of course a degeneracy between $R$ and $a_{\rm min}$. As an
example, one model of the HD~50554 debris disk has $R = 40$ and $a_{\rm
min} = 5.5 \micron$, while another has $R = 45$~AU and $a_{\rm min} = 4.5
\micron$; the difference between their reduced $\chi^2$ statistics is
only 0.2. Yet the value of $R$ from the SED of the resolved source
HD~105211 is in rough agreement with the measured size from the image
(175 AU vs.\ 154 AU), suggesting that we have broken the degeneracy for
at least one source.  Indeed, we find it difficult to reproduce the
measured $32/70 \micron$ flux ratios without setting a minimum grain
size larger than the radiation-pressure blowout size of
\citep{burns79}
\begin{equation}
a_{\rm rp} \geq \frac{6 L_* \langle Q_{\rm pr}(a) \rangle}{16
\pi G M_* c \rho},
\label{eq:radpress}
\end{equation}
where $L_*$ and $M_*$ are the star luminosity and mass, $\langle Q_{\rm
pr}(a) \rangle$ is the radiation pressure coupling coefficient averaged
over all frequencies in the stellar spectrum, and $\rho$ is the particle
density (2.2~g~cm$^{-3}$ for astronomical silicate). In the geometric
optics limit where $\langle Q(a) \rangle = 1$ for all values of $a$, we
find blowout size limits of $0.5-2.2 \micron$ for the range of star
luminosities and spectral types in our sample. Intriguingly,
\citet{pawellek14} and \citet{pawellek15} show that $a_{\rm min} /
a_{\rm rp}$, the ratio of the true minimum grain size to the minimum
predicted value from equation \ref{eq:radpress} (now with $\langle
Q_{\rm pr}(a) \rangle$ calculated self-consistently from grain optical
properties), is a decreasing function of stellar luminosity, approaching
unity for A-type stars but with larger values for Solar-type stars. Our
results agree well with theirs: the most luminous star in our sample,
HD~105211 with $L_*/L_{\odot} > 7$, has $a_{\rm min} / a_{\rm rp} \sim
2$, while our least luminous star, HD~33636 with $L_*/L_{\odot} = 1.05$,
has $a_{\rm min} / a_{\rm rp} \sim 6$.


We suggest that our {\it Herschel} observations most likely trace
dust near the ``birth ring'' \citep{strubbe06}, where the source
planetesimals reside. The largest grains that can emit like blackbodies
at $160 \micron$ have radii set by $2 \pi a \approx \lambda$, such that
$a \approx 25 \micron$. The lifetime $\tau$ of a grain against
Poynting-Robertson drag is
\begin{equation}
\tau = \frac{16 \pi c^2 \rho R^2 a}{3 L_*} ,
\label{eq:prdrag}
\end{equation}
where $c$ is the speed of light. For $25 \micron$ grains, we find $0.2
{\rm Gyr} \leq \tau \leq 2 {\rm Gyr}$ for all of the disks in our
sample. The quantity ${\rm star \; age} / \tau$ for $25 \micron$ grains
is between 0.7 and 6.5 for all systems except one; HD~50554 has age
estimates that differ by a factor of 10 (3.3~Gyr from \citet{bonfanti15}
vs.\ 0.33~Gyr from \citet{chen14}), so ${\rm star \; age} / \tau$ is
either 1.6 or 16. Since the grains that can contribute significantly to
the $160 \micron$ emission have lifetimes against radiative drag of {\it
at least} 1/4 the star age, they are likely to be destroyed by
collisions before migrating significantly. Similarly, all systems have
$\beta < 0.05$ (where $\beta = F_{\rm pr}/F_g$; $F_{\rm pr}$ is the
radiation pressure force given $\langle Q_{\rm pr}(a) \rangle = 1$ and
$F_g$ is the gravitational force) for $25 \micron$ grains, so their
orbits are not substantially modified by radiation pressure. The dust
ring widths from our SED fits are likely close to the true widths of the
source planetesimal belts, with two caveats: (a) mutual grain collisions
will increase the eccentricity dispersion and therefore the ring width
\citep{thebault09}, and (b) our models do not account for the
possibility that the small grains have a wider spatial distribution than
the large grains---we assume all grain sizes are well mixed throughout
the dust annulus.  In reality, radiation pressure might push the
smallest grains to wide or eccentric orbits \citep{burns79, gaspar12}
and create structures similar to the $\beta$~Pictoris, HR~8799, and
HD~95086 ``halos'' seen in scattered light
\citep{augereau01,su09,su15,ballering16}.  Pushing the small grains in
the HD~105211 system to wider orbits might help resolve the conflict
between the $32/70 \micron$ flux ratio and the $160 \micron$ flux
density discussed in section \ref{sec:HD105211}. The HD~105211 disk also
has the smallest $a_{\rm min}/a_{\rm rp}$ size ratio of any disk in the
sample, so is most likely to be affected by radiation pressure.
Follow-up with ground-based adaptive optics or the Hubble Space
Telescope could detect scattered light from blowout-size grains. If our
{\it Herschel} photometry traces the birth rings, our systems can be
added to the list of $> 30$ debris disks with radii larger than the
probable solar nebula radius of 80~AU \citep{kretke12}. Planetesimal
formation seems to be a robust process even out to extremely large
distances \citep{kenyon12}.

Finally, many lines of evidence suggest that at least some planetesimal
belts are sculpted by planets \citep[e.g.][]{quillen02, chiang09,
boley12, su14, sai15}. Interestingly, the disk surrounding HD~50554 is a
true Kuiper Belt analog in its spatial distribution, with a best-fit
radius of 45~AU and ring half-width of 4~AU. Since HD~50554 is slightly
younger than the sun \citep[3.3~Gyr;][]{bonfanti15}, its disk may be
similar to the Kuiper Belt at an earlier stage of evolution and may have
a Neptune-like planet at its inner edge. In N-body simulations of
planetesimal belts with giant planets orbiting just inside their inner
edges, \citet{rodigas14} find that the width of the planetesimal belt
increases with both planet mass and planet/disk semimajor axes. Here we
find that the width of the best-fit dust ring roughly increases with
distance from the star, from $\sigma = 4$~AU at $R = 45$~AU for HD~50554
to $\sigma = 20$~AU for both HD~202206 ($R = 105$~AU) and HD~105211 ($R
= 175$~AU), though the width of the ring becomes less observationally
constrained as $R$ increases.  However, we have no direct constraints on
the presence or absence of planets on wide orbits, and we measure
fractional dust luminosities of $L_{\rm dust}/L_* = 10^{-5} - 10^{-4}$
(Table \ref{tab:seds}), which are consistent with dust production from
icy planetesimals stirred by Pluto-size dwarf planets \citet{kenyon10}.


\section{Conclusions and Future Work}
\label{sec:conclusion}

Here we have presented new {\it Herschel} and updated {\it Spitzer}
photometry of five sunlike stars with debris disks of Kuiper Belt size
and larger. Both extended structure in the images and SED fits from the
Debris Disk Simulator \citep{wolf05} indicate that the disks are not
composed of blackbody grains; instead, small grains are present and
disks are wider than their blackbody radii. As seen by \citet{booth13},
the minimum grain sizes in the best-fit SEDs are in the $3.5 \micron -
4.5 \micron$ size range, larger than the radiation pressure blowout
size. All targets are younger than the sun (though still on the main
sequence), so we may be seeing Kuiper Belt analogs at earlier phases of
evolution. The disks have luminosities consistent with dust production
from icy planetesimals stirred by Pluto-size bodies \citep{kenyon10}.
We also find that the {\it Herschel} $160 \micron$ emission likely
emanates from the birth ring of planetesimals, so we add our targets to
the growing list of Sunlike stars with planetesimals beyond the radius
of the solar nebula.  One caveat is that our SED models may not be
unique: we assumed 100\% astronomical silicate grains and did not
consider mixtures of silicate, ice, and/or organic grains (though icy
grain mantles will likely photoevaporate). Another unexplored
possibility is that different grain sizes have different radial
distributions.

Although three of the stars host planets discovered by Doppler searches
(HD~50554, HD~52265, and HD~202206), the planets are on short-period
orbits that should be dynamically decoupled from the debris disks. Any
planets that sculpt the debris disks in this sample have not yet been
detected. HD~33636 has an M6 companion in a $\sim 3$~AU orbit
\citep{bean07}, indicating that the debris disk at 85~AU is
circumbinary. HD~105211 may also be a spectroscopic binary, though
\citet{eggleton08} find a single star more probable. HD~33636 and
(possibly) HD~105211 complement $\sim 30$ previous detections of
circumbinary debris disks, though there is some indication that the
frequency of cold debris disks may be higher among single stars
\citep{rodriguez15}. There is no conclusive evidence that any of our
targets host ``holey'' debris disks which have planets in the gaps
\citep{kennedy14, meshkat15}, though the IRS spectrum of HD~105211 may
have a peak at $19 \micron$ (see section \ref{sec:HD105211}).

Based on target declination, ALMA follow-up would be possible for
HD~33636, HD~52265, HD~105211, and HD~202206. The most compelling target
is HD~105211 due to both its brightness---0.7~Jy at $100 \micron$---and
its spatial extent of $> 14 \arcsec$, wider than the disk surrounding
the T Tauri star TW~Hydrae. HD~33636 and HD~52265 are also good ALMA
targets based on their extended structure in the {\it Herschel} PACS
images. We also suggest JWST follow-up of HD~105211 using the (a) the
MIRI coronagraph with the $23 \micron$ filter and (b) the MIRI low
resolution spectrometer (LRS). Based on the allWISE W1 photometry, we
find that the the \citet{dodsonrobinson11} photosphere model is likely
accurate, which indicates that HD~105211 has a strong infrared excess at
$23 \micron$.  Observations with JWST would allow us to precisely locate
the inner edge of the cold debris ring and possibly detect a hot inner
debris belt, if it exists.


Funding for this work was provided by NASA research support agreement
1524391. We thank Roberta Paladini of the HASA Herschel Science Center
for guidance on data reduction and Sebastian Wolf for developing the
publicly available Debris Disk Simulator tool. John Gizis and Neal Evans
provided helpful input on data analysis methods. This research has made
use of the following resources: (1) {\it Herschel} Interactive
Processing Environment (HIPE) and the \texttt{ipipe} scripts for data
reduction.  (2) VizieR catalogue access tool, CDS, Strasbourg, France.
The original description of the VizieR service was published in A\&AS
143, 23. (3) NASA/IPAC Infrared Science Archive, which is operated by
the Jet Propulsion Laboratory, California Institute of Technology, under
contract with the National Aeronautics and Space Administration. (4)
SIMBAD database, operated at CDS, Strasbourg, France; \citet{wenger00}.
(5) {\it Herschel} Science Archive,
\texttt{www.cosmos.esa.int/web/herschel/science-archive}. (6)
\texttt{astropy} python library for astronomy, \texttt{www.astropy.org}.
(7) \texttt{APLpy} (Astronomical Plotting Library in Python),
\texttt{aplpy.github.io}.


\begin{deluxetable}{lllrrrrr}
\tabletypesize{\scriptsize}
\tablecaption{Photometry
\label{tab:photometry}}
\tablehead{
\colhead{Instrument} & \colhead{Wavelength} & \colhead{Analysis Method}
& \colhead{HD 33636} & \colhead{HD 50554} & \colhead{HD 52265} &
\colhead{HD 105211} & \colhead{HD 202206}
}
\startdata

MIPS & $24 \micron$ & \citet{chen14} & $43.9 \pm
0.9$\tablenotemark{a,b} &
$51.0 \pm 1.0$\tablenotemark{b} & $76.4 \pm
1.5$\tablenotemark{b} & $367.9 \pm 22.5$\tablenotemark{c} & \nodata \\

MIPS & $24 \micron$ & \citet{su10} & $42.5 \pm 0.5$& $48.6 \pm 0.5$ &
$76.2 \pm 0.8$ & $374.6 \pm 3.8$ & \nodata \\

MIPS & $70 \micron$ & \citet{chen14} & $35.0 \pm
2.7$\tablenotemark{d} & $42.0 \pm 4.5$\tablenotemark{d} &
$38.0 \pm 5.4$\tablenotemark{d} & $473.7 \pm
19.8$\tablenotemark{c} & $28.9 \pm 3.4$\tablenotemark{e} \\

MIPS & $70 \micron$ & \citet{su10} & $38.1 \pm 2.4$ & $43.5 \pm
3.0$ & $40.5 \pm 6.0$ & $702 \pm 38$ & $27.9 \pm 3.4$ \\

\hline

PACS & $70 \micron$ & this work & \nodata & $48.1 \pm 4.3$ &
$40.3 \pm 5.5$ & $692.0 \pm 52.6$ & $33.1 \pm 3.7$ \\

PACS & $100 \micron$ & this work & $42.5 \pm 6.1$ & $39.5 \pm
10.0$ & $37.3 \pm 11.1$ & $704.0 \pm 95.1$ & $43.4 \pm 4.9$ \\

PACS & $160 \micron$ & this work & $30.4 \pm
10.2$ & $23.7 \pm 8.8$\tablenotemark{f} &
undetected & $440.3 \pm 146.0$ & $36.1 \pm 7.5$
\\

\enddata
\tablenotetext{a}{Flux density units are mJy. Photometry has not
been color-corrected.}
\tablenotetext{b}{Original source: IPAC IRSA Spitzer Enhanced
Imaging Products Catalog,
irsa.ipac.caltech.edu/data/SPITZER/Enhanced/Imaging}
\tablenotetext{c}{Original source: \citet{beichman06}}
\tablenotetext{d}{Original source: \citet{trilling08}}
\tablenotetext{e}{Original source: \citet{bryden09}}
\tablenotetext{f}{Detection significance is between $2\sigma$
and $3\sigma$.}
\end{deluxetable}

\begin{deluxetable}{lrcrccrrc}
\tabletypesize{\scriptsize}
\tablecaption{Dust ring parameters from SED fits and images
\label{tab:seds}}
\tablehead{ \colhead{Star} & \colhead{$R_{\rm SED}$
(AU)\tablenotemark{a}} 
& \colhead{$\sigma$ (AU)\tablenotemark{b}} & \colhead{$a_{\rm min}$
($\micron$)\tablenotemark{c}} & \colhead{$M_{\rm
dust} \left(M_{\odot}\right)$\tablenotemark{d}} &
\colhead{$T_{\rm BB}$ (K)\tablenotemark{e}} &
\colhead{$R_{\rm BB}$ (AU)\tablenotemark{f}} &
\colhead{$\Gamma$\tablenotemark{g}} & \colhead{$L_{\rm
dust}/L_*$\tablenotemark{h}}
}
\startdata
HD 33636 & 85 & 12 & 3.0 & $10^{-8}$ & 46 & 37 & 2.3 & $5.4
\times 10^{-5}$ \\
HD 50554 & 45 & 4 & 4.5 & $3 \times 10^{-9}$ & 59 & 27 & 1.7 &
$4.1 \times 10^{-5}$ \\
HD 52265 & 70 & 10 & 3.0 & $3 \times 10^{-9}$ & 58 & 32 & 2.2 &
$2.0 \times 10^{-5}$ \\
HD 105211 & 175 & 20 & 4.1 & $6.6 \times 10^{-8}$ & 49 & 86 &
2.0 & $5.4 \times 10^{-5}$ \\
HD 202206 & 105 & 20 & 3.0 & $4 \times 10^{-8}$ & 45 & 40 & 2.7
& $1.3 \times 10^{-4}$ \\
\enddata
\tablenotetext{a}{Distance of the center of the dust ring from
the star, computed from SED}
\tablenotetext{b}{Half-width of debris ring, assuming Gaussian
number density profile for dust}
\tablenotetext{c}{Minimum grain size}
\tablenotetext{d}{Dust mass}
\tablenotetext{e}{Temperature of dust grains, assuming blackbody
emission}
\tablenotetext{f}{Radius of disk of blackbody grains}
\tablenotetext{g}{$R/R_{\rm BB}$}
\tablenotetext{h}{Ratio of dust luminosity to star luminosity}
\end{deluxetable}



\clearpage
\begin{turnpage}
\begin{deluxetable}{rllcccccl}
\tablewidth{0pt}
\tabletypesize{\tiny}
\tablecaption{Target stars
\label{tab:targets}}
\tablehead{
\colhead{HD} & \colhead{Name} & \colhead{Spectral Type} & \colhead{$V$ 
(mag)\tablenotemark{a}} & \colhead{$d_*$ (pc)\tablenotemark{b}} &
\colhead{$L_*/L_{\odot}$ (this work)} &
\colhead{$L_*/L_{\odot}$ (literature)\tablenotemark{c}} & \colhead{Age (Gyr)\tablenotemark{c}} &
\colhead{Observation IDs\tablenotemark{d}}
}
\startdata
33636 & & G0VH-03 & 7.06 & 28.7 & 1.05 & 1.08 & 2.5 & 134226927[2/3] \\
50554 & & F8V & 6.84 & 31.0 & 1.51 & 1.37 & 3.3 & 134223115[2/3],
134226899[6/7], 134226899[8/9] \\
52265 & & G0V & 6.30 & 28.1 & 2.04 & 2.08 & 2.6 & 134223156[4/5],
134226925[8/9] \\
105211 & $\eta$~Cru & F2V & 4.15 & 19.7 & 7.31 &
7.06\tablenotemark{e} & 1.4\tablenotemark{f} & 134226237[1/2],
134226237[3/4] \\
202206 & & G6V & 8.08 & 46.3 & 1.07 & 1.02 & 1.1 & 134221939[7/8],
134223168[6/7] \\
\enddata
\tablenotetext{a}{ $V$ magnitudes were obtained from color
transformations of the Tycho-2 $V_T$ photometry \citep{hog00}.}
\tablenotetext{b}{ Distances come from the Hipparcos reduction of
\citet{vanleeuwen07a, vanleeuwen07b}.}
\tablenotetext{c}{ Luminosities and ages from \citet{bonfanti15} unless
otherwise indicated.}
\tablenotetext{d}{ Observation IDs recorded in Herschel Science Archive
v7.1.1. Notation: 134226927[2/3] = cross-scan pair of 1342269272 and
1342269273.}
\tablenotetext{e}{ Luminosity from \citet{mcdonald12}.}
\tablenotetext{f}{ Age from \citet{chen14}.}
\end{deluxetable}
\end{turnpage}
\clearpage
\global\pdfpageattr\expandafter{\the
\pdfpageattr/Rotate 90}

\end{document}